\documentclass{article}
\usepackage[utf8]{inputenc}
\usepackage[]{geometry}
\usepackage{comment}
\usepackage{amssymb,amsmath,amsthm,amsfonts}
\usepackage{braket}
\usepackage{fancyhdr}
\usepackage{lastpage}
\usepackage[labelfont=bf]{caption}
\usepackage{etoolbox}
\usepackage{graphicx}
\usepackage {gensymb}
\usepackage{authblk}

\usepackage[sorting=none]{biblatex}
  
\newcommand\mat[1]{\mathcal{#1}}

\title{Different forms of first order spin-orbit motion and their utility in spin matching in electron storage rings\thanks{This work was supported in part by the U.S. Department of Energy, Office of Science, Office of Workforce Development for Teachers and Scientists under the Science Undergraduate Laboratory Internships Program, and by Brookhaven Science Associates, LLC under Contract No. DE-SC0012704 with the U.S. Department of Energy.}}
\author[1]{M. G. Signorelli\thanks{mgs255@cornell.edu}}
\author[1,2]{G. H. Hoffstaetter}
\affil[1]{\textit{Cornell University, Ithaca, NY 14853}}
\affil[2]{\textit{Brookhaven National Laboratory, Upton, NY 11973}}
\date{}

\begin{document}
\maketitle

\begin{abstract}
We derive the first order phase space dependence of spin-orbit motion of a particle in an accelerator by expanding the Thomas-BMT equation. Different forms can be found in the literature and we show how these are related, and care is taken to include fringe fields. The advantages of using certain forms is demonstrated by a detailed re-derivation of the spin matching conditions by V. Ptitsyn for the spin rotators in the Electron Storage Ring (ESR) of the Electron-Ion Collider (EIC) at Brookhaven National Laboratory.
\end{abstract}

\clearpage

\section{First Order Spin-Orbit Motion}
\subsection{Thomas-BMT Equation}
To specify the equation of motion for polarized particles, the right-handed, co-moving orthonormal basis $(\hat{x},\hat{y},\hat{s})$ is used, where $\hat{x}=\hat{x}(s)$ points in the horizontal, $\hat{y}=\hat{y}(s)$ in the vertical, and $\hat{s}=\hat{s}(s)$ in the longitudinal, all at some position $s$ along the trajectory of the particle. 
Whenever first order approximations are made, an ``$\approx$" sign is used. A subscript $0$ specifies a quantity defined on the design orbit, which is assumed to be the zero orbit. We start with the Thomas-BMT equation, which defines the rest-frame spin dynamics given laboratory frame electromagnetic fields and momentum \cite{thomas1927, BMT, thomas1982}. $\vec{S}$ is a 3-vector of the spin expectation values in each direction of the co-moving coordinate system, and $\vec{\kappa}$ is a vector of the path curvatures in the $x$ and $y$ directions \cite{georg}. The term $\vec\kappa\times\hat s$ in $\vec\Omega_t$ accounts for the rotation of the coordinate system. The equation of spin precession has the form
\begin{equation}
\frac{d\vec{S}}{dt} =\vec\Omega_t\times\vec S \ , \ \ \vec\Omega_t = -\frac{q}{\gamma m} \left[(1+G\gamma)\vec{B_\perp} + (1+G)\vec{B_\parallel} + \left(G\gamma+\frac{\gamma}{1+\gamma}\right)\frac{\vec{E}\times\vec{\beta}}{c}\right] -\frac{ds}{dt} \vec{\kappa}\times\hat{s} \ .
\end{equation}

The time derivative of a particle's position $\vec{r}$ gives the particle's velocity, where $'$ denotes differentiation w.r.t. the position along the trajectory $s$,

\begin{equation}
\vec{r} = \vec{r}_0+x\hat{x}+y\hat{y}\ , \ \ 
\vec{v}  = \frac{d\vec{r}}{dt} = \frac{ds}{dt}\left[x'\hat{x}  + y'\hat{y}+ \left(1+\kappa_xx+\kappa_yy\right)\hat{s}\right]
\end{equation}
The phase space coordinates in which we subsequently linearize are the deviations in position and slope from the reference trajectory, $x$, $y$, $x'$, $y'$, and the relative momentum deviation $\delta = \frac{p-p_0}{p_0}$.
The velocity $v$ is
\begin{equation}
v = \frac{ds}{dt}\sqrt{\left(1+\kappa_xx+\kappa_yy\right)^2+x'^2+y'^2} 
\approx \frac{ds}{dt}\left(1+\kappa_xx+\kappa_yy\right) \ .
\end{equation}

Making the substitution $\frac{d\vec{S}}{ds} = \frac{d\vec{S}}{dt}\frac{dt}{ds}$ and $\frac{q}{\gamma m} = \frac{vq}{p}\approx\frac{q}{p}\frac{ds}{dt}\left(1+\kappa_xx+\kappa_yy\right)$, the equation can be rewritten as $\frac{d\vec{S}}{ds} = \vec{\Omega} \times \vec{S}$ with
\begin{equation}
\label{eq:omega1}
\vec{\Omega} \approx -\frac{q}{p}(1+\kappa_x x+\kappa_y y)\left[(1+G\gamma)\vec{B_\perp} + (1+G)\vec{B_\parallel}+\left(G\gamma+\frac{\gamma}{1+\gamma}\right)\frac{\vec{E}\times{\vec\beta}}{c}\right] - \vec{\kappa}\times\hat{s} \ .
\end{equation}

For the magnetic rigidity we linearize $\frac{q}{p} \approx \frac{q}{p_0}\left(1-\delta\right)$. 
Next, expanding $\gamma$ to first order around the design energy gives
\begin{align}
\gamma \approx \gamma_0+\Delta p\left. \frac{d\gamma}{dp}\right\rvert_{p=p_0} = \gamma_0\left(1+\delta\frac{p_0}{\gamma_0}\left. \frac{d\gamma}{dp}\right\rvert_{p=p_0}\right)
=\gamma_0\left(1+\beta_0^2\delta\right)\ ,
\end{align}
where $\gamma = \sqrt{1+(\frac{p}{mc})^2}$ was used. To first order, similarly
\begin{align}
    \gamma\beta= \frac{p}{mc}  \label{eq:gammabeta}
    \approx
    \gamma_0\beta_0\left(1+\delta\right)\ ,
\end{align}



and

\begin{equation}
    \frac{\gamma\beta}{1+\gamma}=\frac{\gamma_0\beta_0(1+\delta)}{1+\gamma_0}(1-\frac{\gamma_0\beta_0^2}{1+\gamma_0}\delta) =\frac{\gamma_0\beta_0}{1+\gamma_0}\left(1+\frac{\delta}{\gamma_0}\right)\ .
\end{equation}

Substituting these values into Eq.~\eqref{eq:omega1}, the spin precession vector to first order is, with the unit vector $\hat \beta$ in the flight direction,

\begin{align}
\vec{\Omega} 
&\approx -\frac{q}{p_0}\Biggl\lbrace\left[\left(1+G\gamma_0\right)\left(1+\kappa_xx+\kappa_yy\right)-\left(1+\frac{G}{\gamma_0}\right)\delta\right]\vec{B}_\perp + \left(1+G\right)\left(1+\kappa_xx+\kappa_yy-\delta\right)\vec{B}_\parallel\Biggr.\nonumber\\
&\quad\quad\quad\quad\quad\Biggl.+\left[\left(G\gamma_0\beta_0+\frac{\gamma_0\beta_0}{1+\gamma_0}\right)\left(1+\kappa_xx+\kappa_yy\right)+\frac{\gamma_0\beta_0}{1+\gamma_0}\left(\frac{1}{\gamma_0}-1\right)\delta\right]\frac{\vec{E}\times\hat{\beta}}{c}\Biggr\rbrace - \vec{\kappa}\times\hat{s}  \ .\label{eq:Omega}
\end{align}

The strengths of the horizontal bending, vertical bending, solenoidal, quadrupole, and skew quadrupole magnetic fields are defined as functions of $s$.

\subsection{Magnetostatic Multipole Fields}

The horizontal bending, vertical bending, solenoidal, quadrupole, and skew quadrupole magnetic field strengths are functions of $s$ defined as

\begin{subequations}
\begin{align}
    \kappa_x&=\frac{q}{p_0}\left.B_y(x,y,s)\right\rvert_{(x,y)=(0,0)} \ ,\\
    \kappa_y&=-\frac{q}{p_0}\left.B_x(x,y,s)\right\rvert_{(x,y)=(0,0)} \ , \\
    K_s &= \frac{q}{p_0}\left.B_s(x,y,s)\right\rvert_{(x,y)=(0,0)} \ ,\\
    K_1 &= \frac{q}{p_0}\left.\partial_y B_x(x,y,s)\right\rvert_{(x,y)=(0,0)} =  \frac{q}{p_0}\left.\partial_x B_y(x,y,s)\right\rvert_{(x,y)=(0,0)} \ , \\
    \tilde{K}_1 &= \frac{q}{2p_0}\left[\partial_x B_x(x,y,s)-\partial_yB_y(x,y,s)\right]_{(x,y)=(0,0)} \ .
\end{align}
\end{subequations}

By Maxwell's equations, $\vec\nabla\times\vec{B}=0$ and $\vec\nabla\cdot\vec{B}=0$. This implies that $\vec{B}=-\vec{\nabla}\psi$ and $\nabla^2\psi=0$ where $\psi=\psi(x,y,s)$ is some scalar potential. Expanding around the design orbit for first order magnetic fields, this gives
\begin{equation}
    \psi\approx\frac{p_0}{q}\left[-\int K_s ds +\frac{x^2+y^2}{4}K_s' - \kappa_xy +\kappa_yx - K_1xy + \frac{\tilde{K}_1}{2}\left(x^2-y^2\right)\right] \ ,
\end{equation}
\begin{align}
    &\vec{B} \approx\nonumber\\ &-\frac{p_0}{q}\left[\left( \frac{1}{2}K_s'x+\kappa_y-K_1y+\tilde{K}_1x\right)\hat{x}+\left( \frac{1}{2}K_s'y-\kappa_x-K_1x-\tilde{K}_1y\right)\hat{y}+\left(-K_s-\kappa_x'y+\kappa_y'x\right)\hat{s}\right] \ .
\end{align}

The parallel field is
\begin{align}\label{eq:B_parallel1}
    \vec{B}_\parallel &= \frac{1}{v^2}\left(\vec{v}\cdot\vec{B} \right)\vec{v} \nonumber \\ 
    &\approx -\frac{p_0}{q}\left[-K_sx'\hat{x}-K_sy'\hat{y}+\left(-K_s + \kappa_yx'-\kappa_xy'- \kappa_x'y+\kappa_y'x\right)\hat{s}\right] \ ,
\end{align}

and thus the perpendicular

\begin{align} \label{eq:B_perp1}
    \vec{B}_\perp &= \vec{B}-\vec{B}_\parallel \nonumber\\
    &\approx -\frac{p_0}{q}\left[\left(\frac{1}{2}K_s'x+\kappa_y-K_1y+\tilde{K}_1x+K_sx' \right)\hat{x} \right.\nonumber \\   
    &\quad\quad\quad\quad\quad \left.+\left( \frac{1}{2}K_s'y - \kappa_x-K_1x-\tilde{K}_1y+K_sy'\right)\hat{y} + \left(\kappa_xy'-\kappa_yx'\right)\hat{s}\right] \ .
\end{align}
Subsituting Eq.~\eqref{eq:B_parallel1} and Eq.~\eqref{eq:B_perp1} into Eq.~\eqref{eq:Omega} gives one form for $\vec{\Omega}$ to first order. The two solenoid and dipole cross terms that appear, $K_s\kappa_xx$ and $K_s\kappa_yy$, are assumed to be zero to first order; that is, when there are overlapping solenoid and dipole fields, one of them is small. Letting $\vec{\Omega}_0$ specify the zero orbit spin precession and $\vec{\omega}$ the perturbative spin precession for first order orbit motion, we obtain
\begin{subequations} \label{eq:omega_0}
\begin{align}
    \Omega_{0x} &\approx G\gamma_0\kappa_y\\
    \Omega_{0y} &\approx -G\gamma_0\kappa_x\\
    \Omega_{0s} &\approx -\left(1+G\right)K_s \ ,
\end{align}
\end{subequations}
\begin{subequations} \label{eq:omega_1}
\begin{align}
    \omega_x &\approx \left(1+G\gamma_0\right)\left(\frac{1}{2}K_s'x+\kappa_y^2y-K_1y+\tilde{K}_1x+\kappa_x\kappa_yx\right)+\left(G\gamma_0-G\right)K_sx'-\left(1+\frac{G}{\gamma_0}\right)\kappa_y\delta\\
    \omega_y &\approx \left(1+G\gamma_0\right)\left(\frac{1}{2}K_s'y-\kappa_x^2x-K_1x-\tilde{K}_1y-\kappa_x\kappa_yy\right)+\left(G\gamma_0-G\right)K_sy'+\left(1+\frac{G}{\gamma_0}\right)\kappa_x\delta\\
    \omega_s &\approx -\left(1+G\right)\left(\kappa_x'y-\kappa_y'x-K_s\delta\right)+\left(G\gamma_0-G\right)\left(\kappa_xy'-\kappa_yx'\right) \ .
\end{align}
\end{subequations}

This form is used in the fast first-order spin tracking method \textit{Sprint}, which is implemented in Bmad using quarternions \cite{jacob}. An equivalent form for $\vec{\omega}$ can be obtained by substituting in the equations of motion, defined by $\frac{d\vec{p}}{dt}=q\vec{v}\times\vec{B}$. The kinetic momentum and its time derivative are
\begin{align}
\vec{p} &\approx p_0\left[x'\hat{x} +   y'\hat{y}+(1+\delta)\hat{s}\right] \ , \\
\frac{d\vec{p}}{dt} \label{p_O1}&\approx \frac{ds}{dt}p_0\left[\left(x''-\kappa_x(1+\delta)\right)\hat{x} +\left(y''-\kappa_y(1+\delta)\right)\hat{y} + \left(\kappa_xx'+\kappa_yy'\right)\hat{s}\right] \ .
\end{align}

To first order, the equations of motion are thus

\begin{align}
   x''-K_sy'-\kappa_x\delta &\approx \frac{1}{2}K_s'y-\kappa_x^2x-K_1x-\tilde{K}_1y-\kappa_x\kappa_yy \ \ , \label{eq:eom1}\\
    -y''-K_sx'+\kappa_y\delta &\approx \frac{1}{2}K_s'x+\kappa_y^2y-K_1y+\tilde{K}_1x+\kappa_x\kappa_yx\ \ . \label{eq:eom2}
\end{align}

Substituting Eq.~\eqref{eq:eom1} and Eq.~\eqref{eq:eom2} into Eq.~\eqref{eq:omega_1} gives a different form of the precession vector,

\begin{subequations} \label{eq:omega_2}
\begin{align}
    \omega_x &\approx -\left(1+G\gamma_0\right)y''-\left(1+G\right)K_sx'+\left(G\gamma_0-\frac{G}{\gamma_0}\right)\kappa_y\delta\\
    \omega_y &\approx \left(1+G\gamma_0\right)x''-\left(1+G\right)K_sy'-\left(G\gamma_0-\frac{G}{\gamma_0}\right)\kappa_x\delta\\
    \omega_s &\approx -\left(1+G\right)\left(\kappa_x'y-\kappa_y'x-K_s\delta\right)+\left(G\gamma_0-G\right)\left(\kappa_xy'-\kappa_yx'\right) \ .
\end{align}
\end{subequations}

In this form of $\vec{\omega}$, it is clear that the spin of a particle moving through perpendicular fields precesses at rate of $(1+G\gamma_0)$ greater than that of the velocity vector precession $x''$ and $y''$. This form proves useful for deriving spin matching conditions, as shown in the next section.

\section{Spin Matching Conditions}

This section entirely follows the derivation and implementation of first order spin matching methods by A. Chao in his SLIM formalism \cite{chao}. The right handed orthonormal coordinate system $(\hat{n}_0, \hat{m}_0, \hat{l}_0)$ is defined, where $\hat{n}_0$ is the 1-turn periodic spin solution on the closed orbit and $(\hat{m}_0,\hat{l}_0)$ precess around $\hat{n}_0$, following spin motion on the closed orbit. The number of precessions of $(\hat{m}_0,\hat{l}_0)$ around $\hat{n}_0$ per turn is the spin tune $\nu_{spin}$. By applying a further rotation to ``wind back" the spin tune instantaneously after each turn, the 1-turn periodic coordinate system $(\hat{n}_0, \hat{m},\hat{l})$ is constructed \cite{deshandbook} where 

\begin{equation} \label{eq:k}
    \hat{m}+\textrm{i}\hat{l} = e^{-\textrm{i}2\pi\nu_{spin}}\left(\hat{m}_0+\textrm{i}\hat{l}_0\right) \ .
\end{equation}

At all other places, $\hat{m}$ and $\hat{l}$ simply rotate with the spin motion on the closed orbit. An individual particle's spin can be expanded around the closed orbit periodic spin direction as

\begin{align} 
    \vec{S}&=\sqrt{1-\alpha^2-\beta^2}\hat{n}_0 + \alpha\hat{m}+\beta\hat{l}\nonumber\\&\approx\hat{n}_0+\alpha\hat{m}+\beta\hat{l}\label{eq:S_SLIM}\ ,
\end{align}

where $\alpha$ and $\beta$ are small. Substitution of Eq.~\eqref{eq:S_SLIM} into $\frac{d\vec{S}}{ds}=\left(\vec{\Omega}_0+\vec{\omega}\right)\times\vec{S}$ gives
\begin{align}
\frac{d}{ds}\left(\hat{n}_0+\alpha\hat{m}+\beta\hat{l}\right) = \left(\vec{\Omega}_0+\vec{\omega}\right) \times \left(\hat{n}_0+\alpha\hat{m}+\beta\hat{l}\right) \ ,
\end{align}

which, to first order in phase space coordinates and $\alpha,\beta$, simplifies to
\begin{align}
    \alpha'\hat{m}+\beta'\hat{l} &\approx \vec\omega\times\hat n_0 \nonumber\\
    &= \vec{\omega}_\perp \times \hat{n}_0 \nonumber\\
    &=\left[(\vec\omega\cdot\hat{l})\hat{l} +(\vec\omega\cdot\hat{m})\hat{m}\right]\times\hat n_0 \nonumber \\
    &= (\vec\omega\cdot\hat{l})\hat{m}-(\vec\omega\cdot\hat{m})\hat{l} \ .
\end{align}

This gives the two differential equations for $\alpha$ and $\beta$,

\begin{subequations}
\begin{align}
    \frac{d\alpha}{ds} &= \vec{\omega}\cdot\hat{l}  \ , \\
     \frac{d\beta}{ds} &= - \vec{\omega}\cdot\hat{m} \ .
\end{align}
\end{subequations}


Letting $\vec{k} = \hat{l}+\textrm{i}\hat{m}$, first order spin matching at an azimuth $s_0$ may be achieved by demanding that the 1-turn spin-orbit integral be equivalent to zero. For reasons that will become apparent shortly, we explicitly specify that $\vec{\omega}=\vec{\omega}(\vec{r})$ where $\vec{r} = (x, p_x, y, p_y, z, \delta)^T$. That is,

\begin{equation} \label{eq:s_o_integral}
\int_{s_{0}}^{s_{0}+C} \vec{\omega}(\vec{r})\cdot\vec{k} \, ds = 0 \ .
\end{equation}

A general $\vec{r}$ can be defined as a linear combination of the orbital eigenvectors $\vec{v}_j$, where $j=\pm I, \pm II, \pm III$ and $\left(\vec{v}_{j}\right)^*=\vec{v}_{-j}$, of the 6D 1-turn transfer matrix. The eigenvectors are normalized so that $\left(\vec{v}_{j}\right)^\dagger\mathbf{S}\vec{v}_{j}= \pm \textrm{i}$ where $\mathbf{S} = \textrm{blkdiag}\left(\mathbf{S}_2,\mathbf{S}_2,\mathbf{S}_2 \right)$ and $\mathbf{S}_2=\begin{pmatrix}
     0 & -1 \\ 1 & 0
 \end{pmatrix}$. In action-angle variables, this is

 \begin{equation}
     \vec{r} = \sum_{j = I,II,III} \sqrt{J_i}\left[\vec{v}_je^{-\textrm{i}\Phi_j} +\vec{v}_{-j}e^{\textrm{i}\Phi_j} \right] \ .
 \end{equation}
 


Spin matching for one of the three modes of oscillation is achieved when the spin orbit integral is zero along that particular eigenmode. For example, in an uncoupled ring where the $I$-mode is the horizontal, a horizontal spin match at azimuth $s_0$ is achieved when

\begin{equation}
\label{eq:s_o_example}
\int_{s_{0}}^{s_{0}+C} \vec{\omega}(\vec{v}_I)\cdot\vec{k} \, ds = \int_{s_{0}}^{s_{0}+C} \vec{\omega}(\vec{v}_{-I})\cdot\vec{k} \, ds = 0 \ .
\end{equation}

We note, by definition, when $\int_{s_{0}}^{s_{0}+C} \vec{\omega}(\vec{v}_j)\cdot\vec{k} \, ds = 0$, that $\int_{s_{0}}^{s_{0}+C} \vec{\omega}(\vec{v}_{-j})\cdot\vec{k} \, ds = 0$. Spin matching should be achieved at as many azimuths $s_0$ as possible. In a perfectly flat ring with only dipoles and regular quadrupoles, a horizontal and longitudinal spin match is automatically achieved at every azimuth: radiation damping makes the vertical beam size approximately zero, and in the $xz$-plane of the ring are purely vertical fields. Therefore, no matter the horizontal or longitudinal amplitude, a particle only sees fields that are aligned with $\hat{n}_0$. In a ring with spin rotators, where $\hat{n}_0$ is rotated out of the vertical, it is not possible to achieve a spin match at every azimuth. 




\subsection{EIC-ESR}
\hspace{0pt}
\begin{figure}[!h]
\centering
\includegraphics{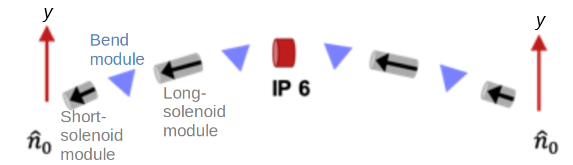}
\caption{Graphic of the spin rotator section in the ESR from \cite{dmarx}.}
\end{figure}

The spin matching conditions for the Electron Storage Ring of the Electron Ion Collider, derived by V. Ptitsyn \cite{Ptitsyn2023,vadimderivation}, are now re-derived in detail to show the utility of the precession vector expressions in Eq.~\eqref{eq:omega_2}. The ESR employs four ``solenoid modules" and ``bend modules" to rotate the spin to longitudinal at the interaction point and back to vertical in the arc. Each solenoid module consists of two equivalent solenoids with some number of quadrupoles in between. These quadrupoles are chosen for decoupling of the solenoid module and for spin matching. The bend modules consist of quadrupoles and horizontal bends. Evaluating the integrand in Eq.~\eqref{eq:s_o_integral} using Eq.~\eqref{eq:omega_2} with the convention for electrons $G\rightarrow a$ gives
\begin{align}
    \vec{\omega}\cdot\vec{k} &= K_s\left[-(1+a)(x'k_{x}+y'k_{y}-\delta k_{s})\right]+\kappa_x\left[(a\gamma_0-a)y'k_{s}-\left(a\gamma_0-\frac{a}{\gamma_0}\right)\delta k_{y}\right] \nonumber\\&\quad\quad\quad\quad\quad- \kappa_x'(1+a)yk_{s}+(1+a\gamma_0)(x''k_{y}-y''k_{x}) \nonumber\\
    &=K_s\left[-(1+a)(x'k_{x}+y'k_{y}-\delta k_{s})\right]+\kappa_x\left[(a\gamma_0-a)y'k_{s}-\left(a\gamma_0-\frac{a}{\gamma_0}\right)\delta k_{y}\right] \nonumber\\&\quad\quad\quad\quad\quad- \kappa_x'(1+a)yk_{s}+(1+a\gamma_0)\left[(x'k_{y})' - x'k_{y}'-(y'k_{x})' + y'k_{x}'\right]\ .
\end{align}

Substitution of $\vec{k}$ into $\frac{d\vec{S}}{ds} = \vec{\Omega}_0\times\vec{S}$ using Eq.~\eqref{eq:omega_0} 
gives

\begin{subequations} \label{eq:k0_eom}
\begin{align}
k_{x}' &= (1+a)K_sk_{y}-a\gamma_0\kappa_xk_{s} \ ,\label{eq:eom_k0x}\\
k_{y}' &= -(1+a)K_sk_{x}\label{eq:eom_k0y} \ ,\\
k_{s}' &= a\gamma_0\kappa_xk_{x} \ .
\end{align}
\end{subequations}


With these equations, the integrand is
\begin{align} \label{eq:s_o_integrand}
        \vec{\omega}\cdot\vec{k} &= K_s(1+a)\left[a\gamma_0\left(x'k_{x}+y'k_{y}\right)+\delta k_{s}\right]-\kappa_x\left[(a+a^2\gamma_0^2)y'k_{s}+\left(a\gamma_0-\frac{a}{\gamma_0}\right)\delta k_{y}\right] \nonumber\\&\quad\quad\quad\quad\quad -\kappa_x'(1+a)yk_{s}+(1+a\gamma_0)\left[(x'k_{y})'-(y'k_{x})'\right] \ .
\end{align}


\subsubsection{Horizontal Spin Match}
We now substitute $\vec{r}=\vec{v}_I$, the horizontal eigenvector, into Eq.~\eqref{eq:s_o_integrand}. With this substitution, the integrals over the $(y'k_{x})'$ and $(x'k_{y})'$ terms are automatically zero: at the start and exit of the spin rotator we demand no transverse coupling, so $v_{I,p_y}(s_{in})=v_{I,p_y}(s_{out})=0$, and $\hat{n}_0$ is vertical, so $k_{y}(s_{in})=k_{y}(s_{out})=0$. Furthermore, because there is no transverse coupling in the bends, $v_{I,y}=v_{I,p_y}=0$ in the bends. Finally, $v_{I,\delta}=0$. This leaves integrals only over the eight solenoids (two per solenoid module). Using the fact that $x'=p_x+\frac{1}{2}K_sy$ and $y' = p_y-\frac{1}{2}K_sx$, this is expressed in terms of the canonical components of $\vec{v}_I$ as 

\begin{align}
    \int_{s_{in}}^{s_{out}} \vec{\omega}(\vec{v}_I)\cdot\vec{k} \, ds = \int_{\textrm{8 sols}} K_s(1+a)a\gamma_0\left[\left(v_{I,p_x}+\frac{1}{2}K_sv_{I,y}\right)k_{x}+\left(v_{I,p_y}-\frac{1}{2}K_sv_{I,x}\right)k_{y}\right] \, ds \ .
\end{align}

Solving the equations of motion in Eq.~\eqref{eq:eom1} and Eq.~\eqref{eq:eom2} for a solenoid shows that $x'$ and $y'$ rotate together at a rate of $K_s$ radians per unit length. Likewise, solving the equations of motion in Eq.~\eqref{eq:eom_k0x} and Eq.~\eqref{eq:eom_k0y} for a solenoid shows that the transverse spin components $k_{x}$ and $k_{y}$ rotate and at a rate of $(1+a)K_s$ radians per unit length. For electrons, $a\approx0.00116$. In the ESR spin matching conditions derivation, it is assumed that $a$ is sufficiently small so that $(1+a)\approx 1$. With this assumption, the spin and velocity precess at approximately the same rate, and so $\left(x'k_{x}+y'k_{y}\right)$ is constant within each solenoid. Letting $L$ be the length of an individual solenoid, this leaves a sum over each of the eight solenoids in the rotator, 

        \begin{equation}
        a\gamma_0\sum_{j=1}^{\textrm{8 sols}} \left(K_sL\right)_j\left[\left(v_{I,p_x}+\frac{1}{2}K_sv_{I,y}\right)k_{x}+\left(v_{I,p_y}-\frac{1}{2}K_sv_{I,x}\right)k_{y}\right]_j  = 0 \ .
\end{equation}

Because the summand is constant within a single solenoid, it can be evaluated at the start, end, or anywhere in between inside that single solenoid. Once again, there are four solenoid modules each consisting of two equal strength solenoids with equal length $L$. It is convenient to express the spin matching conditions in terms of each solenoid module (each consecutive pair of single solenoids), because we can evaluate the summand right at the start of the first solenoid entering the module, and at the end of the second solenoid exiting the module. With the spin precession through the entire $i$-th solenoid module as $\phi_i=2(K_sL)_i(1+a)\approx 2(K_sL)_i$, we express the horizontal spin matching conditions as

\begin{equation}\label{eq:sm_sum}
    a\gamma_0\sum_{i=1}^4 H_i = 0 \ ,
\end{equation}
\begin{align} \label{eq:esr_sm_cond}
    &H_i = \frac{\phi_i}{2}\left\lbrace\left[\left(v_{I,p_x}+\frac{1}{2}K_sv_{I,y}\right)k_{x}+\left(v_{I,p_y}-\frac{1}{2}K_sv_{I,x}\right)k_{y}\right]_{enter}\right.\nonumber\\
        &\quad\quad\quad + \left.\left[\left(v_{I,p_x}+\frac{1}{2}K_sv_{I,y}\right)k_{x}+\left(v_{I,p_y}-\frac{1}{2}K_sv_{I,x}\right)k_{y}\right]_{exit}\right\rbrace_i\ .
\end{align}

$\left(\vec{v}_{I}\right)_{exit}$ can be expressed in terms of $\left(\vec{v}_{I}\right)_{enter}$ multiplied by the orbital transfer matrix over the $i$-th solenoid module, and $(\vec{k})_{exit}$ can be expressed in terms of $(\vec{k})_{enter}$ multiplied by the spin rotation matrix over the solenoid module. Furthermore, there is no coupling entering and exiting each solenoid module, and so $\left(v_{I,y}\right)_{enter}=\left(v_{I,p_y}\right)_{enter}=\left(v_{I,y}\right)_{exit}=\left(v_{I,p_y}\right)_{exit}=0$.  Explicitly, 

\begin{align}
    (v_{I,x})_{exit}  &= \ \ (v_{I,x})_{enter}m_{11} \ +(v_{I,p_x})_{enter}m_{12} \ , \\
    (v_{I,p_x})_{exit} &= \ \ (v_{I,x})_{enter}m_{21}\ +(v_{I,p_x})_{enter}m_{22} \ , \\
    (k_x)_{exit} &= \ \ (k_x)_{enter}\cos{\phi_i} + (k_y)_{enter}\sin{\phi_i}\ , \\ 
    (k_y)_{exit} &= -(k_x)_{enter}\sin{\phi_i} + (k_y)_{enter}\cos{\phi_i} \ . 
\end{align}

Thus,


\begin{align} \label{eq:H}
    H_i &= \frac{\phi_i}{2}\left\lbrace v_{I,p_x}\left[k_x\left(1+m_{22}\cos{\phi_i} + m_{12}\frac{1}{2}K_s\sin{\phi_i}\right) + k_y\left( m_{22}\sin{\phi_i} - m_{12}\frac{1}{2}K_s\cos{\phi_i}\right) \right] \right. \nonumber\\
    &\left. v_{I,x}\left[k_x\left(m_{21}\cos{\phi_i}+m_{11}\frac{1}{2}K_s\sin{\phi_i}\right) + k_y\left(-\frac{1}{2}K_s + m_{21}\sin{\phi_i} - m_{11}\frac{1}{2}K_s \cos{\phi_i}\right)\right] \right\rbrace_{enter} \ .
\end{align}

The horizontal spin matching conditions can be satisfied by setting $H_i$ for each solenoid module such that the sum in Eq.~\eqref{eq:sm_sum} is equal to zero. One way of doing this is setting each $H_i=0$. We also seek a solution which does not depend on the incoming orbital eigenvector. The goal therefore is to choose the transfer matrix components over the solenoid module so that the term proportional to $v_{I,x}$ and that proportional to $v_{I,p_x}$ are each zero. First, we consider the solution which is independent of the incoming $\vec{k}$, such that each term proportional to the components of $\vec{k}$ is zero. Solving the system for $m_{22}$ and $m_{12}$ (proportional to $v_{I,p_x}$ and for $m_{21}$ and $m_{11}$ (proportional to $v_{I,x}$), and taking care to never divide by $\cos{\phi_i}$ which will equal zero when $\phi_i = 90\degree$, gives the transfer matrix elements for an arbitrary incoming $\vec{k}$, the so-called \textit{universal solution},

\begin{align}
\mat{M}_{i,x} = 
\begin{pmatrix}
-\cos{\phi_i} &  -\frac{2}{K_s}\sin{\phi_i}\\
\frac{K_s}{2}\sin{\phi_i} & -\cos{\phi_i}
\end{pmatrix} \ .\end{align}

It seems, by coincidence, that the universal solution satisfies the symplectic condition $\det{(\mat{M})}=1$. Now, we consider possible other solutions dependent on the incoming $\vec{k}$. From this point forward, we will omit the subscript $enter$ and assume it is implied. We rewrite Eq.~\eqref{eq:H} as

\begin{align}
    H_i &= \frac{\phi_i}{2}\left\lbrace v_{I,p_x}\left[m_{22}\left(k_x\cos{\phi_i} +k_y\sin{\phi_i}\right)+ m_{12}\left(k_x\frac{1}{2}K_s\sin{\phi_i} - k_y\frac{1}{2}K_s\cos{\phi_i}\right) + k_x \right] \right. \nonumber\\
    &\quad\quad\left. v_{I,x}\left[m_{21}\left(k_x\cos{\phi_i} +k_y\sin{\phi_i}\right)+ m_{11}\left(k_x\frac{1}{2}K_s\sin{\phi_i} - k_y\frac{1}{2}K_s\cos{\phi_i}\right) -\frac{1}{2}K_sk_y\right] \right\rbrace \ .
\end{align}

Let 
\begin{align}
    A &= k_x\cos{\phi_i}+k_y\sin{\phi_i} \ ,\\
    B &= k_x\frac{1}{2}K_s\sin{\phi_i} - k_y\frac{1}{2}K_s\cos{\phi_i} \ \ .
\end{align}

The spin matching conditions for any incoming $\vec{k}$ can be written as
\begin{align}
      Am_{22} + Bm_{12} &= -k_x  \ , \label{eq:m22m12} \\
      Am_{21}+Bm_{11} &= \frac{1}{2}K_sk_y \ . \label{eq:m21m11}
\end{align}

Including the symplectic condition, we calculate $m_{11}\times$Eq.~\eqref{eq:m22m12} $-$ $m_{12}\times$Eq.~\eqref{eq:m21m11} to be
\begin{align}
    &A\left(m_{11}m_{22}-m_{12}m_{21}\right)+B\left(m_{12}m_{11}-m_{12}m_{11}\right) = -k_xm_{11}-\frac{1}{2}K_sk_ym_{12} \ ,\nonumber\\
   &\quad\quad\quad\quad\quad\quad\quad\quad\quad\rightarrow A=-k_xm_{11}-\frac{1}{2}K_sk_ym_{12}  \ .
\end{align}
Writing out the full form of $A$, and expressing this in terms of the real and imaginary parts of $\vec{k}$ (defined previously as $\vec{k}=\hat{l}+\textrm{i}\hat{m}$), this is

\begin{align}
-
    \begin{pmatrix}
        l_x\cos{\phi_i}+l_y\sin{\phi_i} \\
        m_x\cos{\phi_i}+m_y\sin{\phi_i}
    \end{pmatrix} = \begin{pmatrix}
        l_x & \frac{1}{2}K_sl_y \\
        m_x & \frac{1}{2}K_sm_y
    \end{pmatrix}
    \begin{pmatrix}
        m_{11}\\m_{12}
    \end{pmatrix} \ . \label{eq:solA}
\end{align}

Similarly, with $m_{22}\times$Eq.~\eqref{eq:m21m11} $-$ $m_{21}\times$Eq.~\eqref{eq:m22m12}, 
\begin{align}\frac{1}{2}K_s
\begin{pmatrix}
            l_x\sin{\phi_i} - l_y\cos{\phi_i}\\
            m_x\sin{\phi_i} - m_y\cos{\phi_i}
    \end{pmatrix} = \begin{pmatrix}
        l_x & \frac{1}{2}K_sl_y \\
        m_x & \frac{1}{2}K_sm_y
    \end{pmatrix}
    \begin{pmatrix}
        m_{21}\\m_{22}
    \end{pmatrix} \ .\label{eq:solB}
\end{align}

Solving Eqs.~\eqref{eq:solA} and \eqref{eq:solB} gives the symplectic transfer matrix components that satisfy the horizontal spin matching conditions. Notably, infinitely many solutions will exist when the projections of $\hat{l}$ and $\hat{m}$ on the $xy$-plane lie along the same line (so the rows of the matrix are not linearly independent). This is \textit{always} the case for the first and last solenoid modules, where the entering $\hat{n}_0$ lies entirely in the $xy$-plane. In every other case, the universal solution is the only solution.
\subsubsection{Longitudinal Spin Match}
The periodic dispersion is expressed in terms of the longitudinal eigenvector as $\eta_x=\frac{1}{2}\left(\frac{v_{III,x}}{v_{III,\delta}} + \frac{v_{III,x}^*}{v_{III,\delta}^*}\right)$, $\eta_{p_x}=\frac{1}{2}\left(\frac{v_{III,p_x}}{v_{III,\delta}} + \frac{v_{III,p_x}^*}{v_{III,\delta}^*}\right)$, and likewise for the vertical. Because the spin-orbit integral will by definition be zero for $\left(\vec{v}_{III}\right)^*$ when it is zero for $\vec{v}_{III}$, the notation can be simplified by expressing the longitudinal spin matching conditions in terms of a particle on the dispersive orbit $\vec{r}=\vec{\eta}=(\eta_x,\eta_{p_x},\eta_y,\eta_{p_y},0,1)^T$. We substitute $\vec{\eta}$ into Eq.~\eqref{eq:s_o_integrand}, and evaluate. As with the horizontal spin matching conditions, integrals over the terms $(y'k_{x})'$ and $(x'k_{y})'$ are again zero; at the start and exit of the rotator there is no vertical dispersion and $\hat{n}_0$ is vertical. Furthermore, terms proportional to $y, y'$ in the bends will be zero because there is no vertical dispersion. This leaves
\begin{align}
        \int_{s_{in}}^{s_{out}} \vec{\omega}(\vec{\eta})\cdot\vec{k} \, ds &= \int_{\textrm{8 sols}} K_s(1+a)\left\lbrace a\gamma_0\left[\left(\eta_{p_x}+\frac{1}{2}K_s\eta_y\right)k_{x}+\left(\eta_{p_y}-\frac{1}{2}K_s\eta_x\right)k_{y}\right] +k_s\right\rbrace\, ds \nonumber\\
        &\quad\quad\quad\quad\quad\quad\quad\quad\quad\quad-\int_{\textrm{4 bends}} \kappa_x\left(a\gamma_0-\frac{a}{\gamma_0}\right)k_{y} \, ds \ .
\end{align}

Assuming $(1+a)\approx 1$ so that $(x'k_x +y'k_y)$ is constant within a solenoid, the terms proportional to $a\gamma_0$ in the integral over the solenoids gives the exact same conditions as the horizontal spin matching conditions. This is because there is zero vertical dispersion and zero coupling entering and exiting the solenoid module, and the horizontal dispersion will propagate over the solenoid module according to the same $2\times 2$ horizontal transfer matrix. Defining $\phi_i = 2(K_sL)_i(1+a)\approx 2(K_sL)_i$ as the spin precession in the $i$-th solenoid module, $\psi_j = a\gamma_0(\kappa_xL)_j$ the spin precession in the $j$-th bend module, and finally assuming $a/\gamma_0\approx 0$, the longitudinal spin matching conditions are

\begin{equation}
    a\gamma_0\sum_{i=1}^4 H_i+\sum_{i=1}^4 k_{s,i}\phi_i - \sum_{j=1}^4 k_{y,j}\psi_j = 0 \ .
\end{equation}

Therefore, once horizontally spin-matched, a longitudinal spin match can be achieved entirely by the geometry and spin rotation angles in the bends and solenoids of the rotator. One example of such a configuration is the eRHIC Zeroth Order Design, with an antisymmetric bend arrangement and opposite polarity solenoid modules on either sides of the interaction point \cite{zdr}.

\section{Conclusions}
By expanding the Thomas-BMT equation around small phase space coordinates, the first order spin-orbit motion of a particle in an accelerator including fringe fields was derived. We showed how different forms of the spin precession vector found in the literature are equivalent, related by substitution of the equations of motion. The utility of the different forms was also discussed; the form in Eq.~\eqref{eq:omega_1} proves useful for fast first-order spin tracking, and the form in Eq.~\eqref{eq:omega_2} both elegantly shows the $(1+G\gamma_0)$ amplified spin rotation with velocity vector rotation and is useful for deriving the spin matching conditions in the ESR of the EIC.

\nocite{*}
\printbibliography

\end{document}